\documentclass[showpacs,preprintnumbers,amsmath,amssymb,prb]{revtex4}

\usepackage{graphicx}
\usepackage{dcolumn}
\usepackage{bm}

\begin{document}

\title{Magneto-elastic interaction in cubic helimagnets with B20 structure}

 \author{S. V. Maleyev}
\affiliation{Petersburg Nuclear Physics Institute, Gatchina, St.\ Petersburg 188300, Russia}

\date{\today}
\newcommand{\m}{\mathbf}

\begin{abstract}
The magneto-elastic interaction in cubic helimagnets with B20 symmetry  is considered. It is shown that this interaction is responsible for negative contribution to the square of the spin-wave gap $\Delta$ which is alone has to disrupt assumed helical structure.
It is suggested that competition between positive part of $\Delta^2_I$ which stems from magnon-magnon interaction and its negative magneto-elastic part leads to the quantum phase transition  observed at high pressure in $Mn Si$ and $Fe Ge$. This transition has to occur when $\Delta^2=0$.
For $Mn Si$ from rough estimations at ambient pressure both parts  $\Delta_I$ and  $|\Delta_{ME}|$ are comparable with the experimentally observed gap.
The magneto-elastic interaction is responsible also for $2\m k$ modulation of the lattice where $\m k$ is the helix wave-vector and  contribution to the magnetic anisotropy.

 Experimental observation by $x$-ray  and neutron scattering  the lattice modulation allows determine the strength of anisotropic part of the magneto-elastic interaction  responsible for above phenomena and  the lattice helicity. 
  \end{abstract}


\maketitle
\section{Introduction}

Helical magnetic order stipulated by Dzyaloshinskii-Moriya interaction and its evolution in magnetic field attract a lot of attention in last years (see for example \cite{HK} and references therein).  In this respect
non-centrosymmetric cubic helimagnets such as $Mn Si, Fe Ge, Fe Co Si$ with $P2_13\:(B20)$ symmetry play special role. They  are a subject of the intensive experimental and theoretical studies during  several decades. Their single-handed helical structure was explained by Dzyaloshinskii \cite{D}. The full set of interactions responsible for observed helical structure (Bak-Jensen model) was established later in \cite{N}, \cite{B} in agreement with existing experimental data (see for example \cite{I} and references therein). The renascence in this field began with a discovery of the quantum phase transition to a disordered (partially ordered) state in $Mn Si$  at high pressure (magnetization and resistivity measurements \cite{P1}, \cite{K} and neutron scattering \cite{P2},
\cite{P4}).  Recently similar transition was observed in $Fe Ge$ \cite{PE}. The following properties of this state attract the main attention: i) non-Fermi-liquid conductivity, ii) spherical neutron scattering surface with the weak maxima along the $\langle 1 10\rangle$ axes \cite{P2}, \cite{P3}, whereas at ambient pressure Bragg reflections  and critical scattering were observed along  $\langle 111 \rangle$ \cite{I}. These features and the structure of the partially ordered state were discussed in several theoretical papers (see \cite{T,R,BK,Bn} and references therein). It should be noted  that the spherical scattering surface with maxima along $\langle 111 \rangle$ was observed at ambient pressure just above critical temperature $T_c\simeq 29K$ and explained using the Bak-Jensen model \cite{G1}. It was demonstrated also that this phase transition is of the first order one \cite{ST}   Recently phase separation in $Mn Si$ near the quantum phase transition was observed by  muon spin resonant method \cite{U}.

In \cite{M} \cite{BK} strong anisotropy of the spin-wave spectrum at low momenta was demonstrated: excitations with momentum along and perpendicular to the helix wave-vector $\m k$ have linear and quadratic dispersion respectively. At the same time there is a contradiction between these papers. In \cite{BK} was claimed that the spin-waves are gapless Goldstone excitations due to translation invariance along the helix axis whereas in \cite{M} the spin-wave gap was calculated in $1/S$ approximation. This contradiction was discussed in \cite{M1}. In brief its essence is following. In \cite{BK} the $1/S$ corrections to the spin-wave energy were not evaluated and the translation invariance was not proved. Meanwhile doing this the authors should meet a problem how to consider the Dzyaloshinskii-Moriya interaction. It contains two spin operators and if they belong to the single lattice point, the translation invariance holds and the gap is zero. However this interaction always acts between different spins, the translation changes the pair energy and the gap is not zero  \cite{com}.

Existence of the gap $\Delta$ is very important for correct description of the helix behavior in magnetic field $H_\perp$ perpendicular to the helix vector $\m k$. In the gapless case the spin-wave spectrum becomes  unstable in infinitesimal $H_\perp$ in contradiction to the well-known experimental findings \cite{M} and predictions of the phenomenological Landau-like theory \cite{PW1}.   In \cite{M} was shown  that the helix state remains stable if $H_\perp<\Delta\sqrt 2$ and then $\m k$ begins rotate toward the field. Recently this prediction was confirmed using small angle polarized neutron scattering and was found that  $\Delta\simeq 13\mu e V$ for $Mn Si$ \cite{G3,G2,G4,G5}.

The quantum phase transition to magnetically disordered state observed in $Mn Si$  and $Fe Ge$ \cite{K,P2,P3, PE,P4} 
is an another important   problem. Recently it was considered  on the base of phenomenological Landau theory \cite{GE}. However up to now we have not any attempt to understand microscopic origin of this transition. In this paper we consider the magneto-elastic (ME) interaction and evaluate its contribution to the square of the spin-wave gap $\Delta$. We demonstrate that 
\begin{eqnarray}
	\Delta^2=\Delta^2_I+\Delta^2_{ME},
\end{eqnarray}
where $\Delta^2_I$ was evaluated in \cite{M} and $\Delta^2_{ME}<0$ appears due to the ME  considered as the second order perturbation. It is important to note that if $\Delta^2_I$ is zero the ME has to disrupt the helical magnetic order.

  Rough estimations  using existing experimental data at ambient pressure (see Sec.V) give $|\Delta_{ME}|\sim 7.6\mu e V$ and $\Delta_I=4.0\div 28 \mu e V$.   Both contributions are comparable with  experimental value determined in \cite{G3}. Hence at pressure  two parts of $\Delta^2$ has to compete and the quantum phase  transition to the partly disordered state occurs when $\Delta=0$. Besides we estimated the ME contribution to the  magnetic anisotropy and demonstrated that it is not small in comparison to the experimental value.  We demonstrated also that the ME leads to the lattice deformation with the wave-vector $2\m k$ and evaluated intensities of corresponding super-lattice reflections. Their experimental study would allow determine
the strength of anisotropic part of the ME interaction responsible for above mentioned phenomena and  the lattice helicity. It has to be noted that the ME interaction in the Landau theory was investigated in \cite{PW, P} and the lattice deformation was predicted at the wave-vector $\m k$. This result does not contradict to ours as it was obtained for magnetized systems only.

The paper is organized as follows. In Sec.II we consider  the ME in cubic helimagnets. Classical ground state energy and the  lattice deformation is studied in Sec.III. The spin-wave-phonon interaction and the ME contribution to $\Delta^2$ is considered in Sec.IV. Obtained results, numerical estimations and experimental consequences are discussed in Sec.V and main results are summarized in Sec.IV In Appendixes A and B some mathematical details are considered. Appendix C is devoted to consideration super-lattice reflections  near forbidden $<n00>$ Bragg peaks with odd $n$-s.

\section{Magneto-elastic interaction}

In general form the  magneto elastic energy is given by (see for example \cite{L})
\begin{equation}
	V_{ME}=\sum_{\m R} S^\alpha_\m R S^\beta_\m R B_{\alpha \beta\gamma\mu}U_{\gamma \mu}(\m R)
\end{equation}
where $S^\alpha_\m R$ is the spin components at the lattice point $\m R$ , $U_{\gamma \mu}=(1/2)(\partial u_\gamma/\partial R_\mu +\partial u_\mu/\partial R_\gamma)$ is the deformation tensor and the lattice site displacement has well known form
\begin{equation}
	\m u(\m R)=\sum e^{i\m{q\cdot R}}\frac{1}{\sqrt{2 N M\omega_{\m q j}}}(\m e_{\m q j}b_{\m q j}+\m e_{-\m q j}b^+_\m {-\m q j}),
\end{equation}
where $\m e_{\m q j}$ are vectors of the phonon polarization , $b(b^+)$ their absorption (excitation) operators and $e_{-\m q j}=e^*_{\m q j}$ \cite{LL}.
  
  Tensor $B$  is symmetric in $(\alpha \beta)$ and $(\gamma \mu)$ components.   In  cubic crystals we have following non-zero components \cite{L}
\begin{equation}
 B_{x x x x}=B_{y y y y}=B_{z z z z}=B_1,\,B_{x y x y}=B_{y z y z}=B_{z x z x}=B_2,
 \end{equation}
 and $B_{x y x y}=B_{y x x y}=B_{x y y x}$ etc. In isotropic medium we have
\begin{equation}
	B^{is}_{\alpha \beta\mu\nu}=B_{is}(\delta_{\alpha\mu}\delta_{\beta\nu}+\delta_{\alpha\nu}\delta_{\beta\mu})/2,
\end{equation}
and $B_1=B_{is}$, $B_2=B_{is}/2$.
 
 In  cubic helimagnets  the lattice spin is given by \cite{M}
\begin{equation}
\begin{aligned}
	\m S_\m R&=\m A e^{i\m{k\cdot R}}(S^\zeta_\m R\cos\alpha+i S^\eta_\m R-S^\xi_\m R\sin\alpha)+\m A^*e^{-i\m{k\cdot R}}(S^\zeta_\m R\cos\alpha-i S^\eta_\m R-S^\xi_\m R\sin\alpha)+\hat c(S^\zeta_\m R\sin\alpha+S^\xi_\m R\cos\alpha)\\&=\m A S^{\m A}_\m R+\m A^* S^{A^*}_\m R+\hat c S^c_\m R,
\end{aligned}	
\end{equation}
where $<S^\zeta_\m R>\neq 0$ is an average value of the lattice spin, $S^{\eta,\xi}_\m R$ describe its perpendicular motion, $\m k$ is the helix wave-vector,  $\m A=(\hat a-i\hat b)/2$, unit vectors $\hat a,\, \hat b,\:\mbox{and}\:\hat c$ form right-handed orthogonal frame,  $\sin\alpha=-H_\parallel/H_C$ where $H_\parallel$ is the magnetic field component along the helix vector $\m k$ and  $H_C$ is the critical field for transition to ferromagnetic state. According to \cite{M} the vector $\m k\parallel \hat c$ in arbitrary field.

 Using standard definition $\m S_\m q=N^{-1/2}\sum \m S_\m R\exp{(-i\m{q\cdot \m R)}}$  in momentum space we obtain \cite{M}
 \begin{equation}
	\begin{aligned}
	\m{S_q}&=S^c_\m q \hat c+S^A_\m q \m A+S_\m q^{A^*} \m A^*\\ 
	S^c_\m q&=S^\zeta_\m q  \sin\alpha+S^\xi _\m q  \cos\alpha,\:
	S^A_\m q=S^\zeta_\m{q-k}\cos\alpha-S^\xi_\m{q-k}\sin\alpha+i S^\eta_\m{q-k},\:
	S^{A^*} _\m q=S^\zeta_\m{q+k}\cos\alpha-S^\xi_\m{q+k}\sin\alpha-i S^\eta_\m{q+k},
	\end{aligned}
\end{equation}
where $S^c_\m q,\: S^{\m A}_\m q\:\mbox{and}\:S^{\m A^*}_\m q$ are functions of $ \m q,\: \m {q-k}\:\mbox{and}\:\m{q+k}$ respectively.

The spin components in Eq.(7) have  well-known form
\begin{equation}
	S^\zeta_\m q=N^{1/2}S\delta_{\m q,0}-(a^+a)_\m q;\:S^\eta_\m q=-i\sqrt{S/2}[a_\m q-a^+_{-\m q}-(a^+a^2)_\m q/2S];\:S^\xi=\sqrt{S/2}[a_\m q+a^+_{-\m q}-(a^+a^2)\m q/2S],
\end{equation}
where ~$a_\m q \:\mbox{and}\:a^+_\m q$ ~are conventional  spin-wave  operators.  

In momentum space Eq.(2) is given by
\begin{equation}
	V_{ME}=N^{-1/2}\sum_{l,m=\m A,\m A^*, c}S^l_{\m q_1} S^m_{\m q_2} B U_\m {-q_1-q_2},
\end{equation}
 This expression is divided on three parts: direct  ($\hat c\hat c\:\mbox{and}\,\m{A\,A^*}$) terms where the $U$ tensor is  $\m k$ independent,  first  $(c A),\,(c\m A^*)$ and second ($\m{A\,A}\:\mbox{and}\,{\m A^*\,\m A^*}$) order umklapp terms where $U$ operator depends on $\m{q_1+q_2\pm k}$ and $\m{q_1+q_2\pm 2k}$ respectively \cite{Ref}. For their consideration we use following identity 
\begin{eqnarray}
	m n B U_\m Q=i B_A\sum_{p=x,y,z}m_p n_p Q_p u_p+i B_2\m{[({m\cdot Q})(n\cdot u)+(m\cdot u)(n\cdot Q)]},
\end{eqnarray}
where $B_A=B_1-2B_2$ is an anisotropic part of the tensor $B$ and $\m{u=u_Q}$. 
 
 In the case of uniform pressure $P$ we have $U_{\alpha\beta}=-(P/3K)\delta_{\alpha\beta}$ where $K$ is the bulk modulus and $U_\m{q}\sim N^{1/2}\delta_{\m q, 0}$.  As a result the umklapp terms are zero as $\hat c\cdot \m A=\m{A\cdot A}=0$ and $V_{ME}\to -N B_1 S(S+1)P/(3K)$. Hence the uniform pressure contributes to the classical part of the magneto-elastic  ground state energy only. However it has to change basic parameters of the problem such as $B_{1,2}$, sound velocities etc.
 
 \section{Ground state energy and lattice deformation }
 
In zero magnetic field  $\sin\alpha=0$ and we have a planar helix. In this case  from Eqs.(7-10) follows that in the classical part of the ME interaction the first-order umklapps are forbidden \cite{NB}
 and    we obtain
\begin{eqnarray}
			V_{ME}=-2i N^{1/2}S^2B_A k[(\m{g\cdot u})-c.c],
		\end{eqnarray}
	where	$g_p=A^2_p\hat c_p$and $u_p=u^p_{-2\m k}$.

For evaluation of the lattice deformation  we must consider the elastic energy.   Unlike Refs. \cite{PW, P} for simplicity we ignore in it the cubic symmetry as its principal symmetry breaking   role has been taken into account above in Eq(11). In this case the unit cell  energy is given by \cite{LL1}
\begin{eqnarray}
	F(\m r)=Q[U_{\alpha\beta}(\m r)U_{\beta\alpha}(\m r)+\sigma U^2_{\alpha\alpha}(\m r)/(1-2\sigma)],
\end{eqnarray}
where $Q=E v_0/[2(1+\sigma)]$, $v_0$ is the unit cell volume,  and
 $E$ and $\sigma$ are Young modulus and Poisson coefficient respectively. As a result we have 
\begin{eqnarray}
	F_{2\m k}=2 Q[k^2(\m u\cdot \m u^*)+(\m k\cdot\m u)(\m k\cdot\m u^*)/(1-2\sigma)].
\end{eqnarray}

From Eqs.(11-13) for the ME part of the ground state  energy we obtain
\begin{eqnarray}
E_{ME}=(2N S^4B^2_A/Q)\{(\m w\cdot\m w^*)+(\m w\cdot\hat c)(\m w^*\cdot\hat c)/(1-2\sigma)-[(\m g\cdot\m w)+c.c.]\},
\end{eqnarray}
where $\m u_{-2\m k}=-i(B_A S^2 N^{1/2} /Q k)\m w$.
Minimum of this energy is evaluated in Appendix A and we have
\begin{eqnarray}
	E_{ME}&=&-\frac{N S^4B^2_A(1+\sigma)}{4E v_0}\left[(G_1-G_2)+\frac{(1-2\sigma)}{2(1-\sigma)}G_2\right]\simeq-\frac{N E v_0g^2_A}{4}(G_1-G_2+G_2/2),\\ 
	\m u_{-2\m k}&=&\frac{-2i N^{1/2}S^2B_A(1+\sigma)}{E v_0 k}\left[\m g^* -\frac{(\m g^*\cdot\hat c)\hat c}{2(1-\sigma)}\right]\simeq -(2i g_A/k)[\m g^*-(\m g^*\cdot\hat c)\hat c/2],
\end{eqnarray} 
where $g_A=S^2B_A/E v_0$,  cubic invariants  $G_1=16(\m g\cdot\m g^*)$ and  $G_2=16(\m g\cdot\hat c)(\m g^*\cdot\hat c)$ are considered in Appendix A and in the right hand side of both equations we neglect $\sigma$ as it is usually  small \cite{LL1}.  
 
 \section{magnon-phonon interaction}
 
 We consider now the magnon-phonon interaction. We are interested  by terms which survive at $\m q=0$ and contribute to the spin-wave gap as other terms are small corrections to the $q$-dependent part of the magnon dispersion considered in \cite{M} and \cite{BK}. 
To single out them we have to replace   in  Eq.(9) one of $S^\zeta$ operators by $S\delta_{\m q\pm\m k,0}$. As a result we obtain terms with  phonon   momenta $\m{q,\,q\pm k},\,\mbox{and}\,\m q\pm 2\m k$. The former disappear at $\m q=0$, the second are proportional to $a_\m q+a^+_{-\m q}$ and can not contribute to the gap (see below). Using identity (10) for the last $A A$ end $A^*A^*$ terms in the case of planar helix we have
\begin{eqnarray}
	V_{2\m k}=-(2S)^{3/2}i k B_A\sum[(a_\m q-a^+_{-\m q})g_p u^p_{-2\m k-\m q}+(a_{-\m q}-a^+_\m q)g^*_p u^p_{2\m k+\m q}],
\end{eqnarray}
  From this equation for the magnon-magnon interaction  we obtain
\begin{eqnarray}
	V_{MM}=2(2k S^2B_A)^2/S\sum(a_\m q-a^+_{-\m q})\left[\sum g_p D_{p\, r}(\Delta ,2\m k)g^*_r\right](a_{-\m q}-a^+_\m q),
\end{eqnarray}
 where we neglect $\m q$ in comparison with $\pm 2\m k$ . 
   The phonon Green function
 can be represented as
 \begin{eqnarray}
  D_{p\,r}(\omega,\m Q)=D_t(\delta_{p\,r}-\hat Q_p  \hat Q_r) +D_l\hat Q_p\hat Q_r,
  \end{eqnarray}
  where $\hat Q=\m Q/Q$ and
    $D_{l(t)}=[M(\omega^2-s^2_{l(t)}q^2)]^{-1}$ where $l(t)$labels  longitudinal ( transverse) phonon  mode and $s_{l(t)}$ is a corresponding sound velocity.  We neglect optical branches  as their contribution is of order $(s k/\theta_D)^2\ll 1$ where $\theta_D$ is Debye temperature.
 
   In the linear spin-wave theory the Hamiltonian is given by \cite{M} 
\begin{equation} 
 H_{SW}=\sum[ E_\m q a^+_\m q a_\m q+ B_\m q(a_\m q a_{-\m q}+a^+_{-\m q}a^+_\m q)/2]
 \end{equation}
 and the square of the spin-wave energy  $\epsilon^2_\m q=E^2_\m q-B_\m q^2$. As was shown in \cite{M} $E_0=B_0=A k^2/2$ where $A$ is the spin-wave stiffness at $q\gg k$ and we have gapless excitations. We assume that the ME  interaction is weak  
  and gives small corrections to $A_0$ and $B_0$. In this case from Eqs.(18-20) for the magneto-elastic contribution to the square of the spin-wave gap we obtain
\begin{equation}
	\Delta^2_{ME}=-\frac{A k^2(B_A S^2)^2}{2S M}\left(\frac{G_1-G_2}{s^2_t}+\frac{G_2}{s^2_l}\right)\simeq-\frac{A k^2E v_0 g_A^2}{4S}(G_1+G_2)
\end{equation}
where $g_A=S^2B_A/E v_0$ and neglecting $\sigma$ we have $M s^2_l=E v_0$ and $s_t=s_l/\sqrt 2$.
This expression is negative  as it should be in the second order perturbation theory and $\Delta^2_{ME}=0$ in $<100>$ direction only (see Appendix A). Consideration of $\m{q\pm k}$ terms lead to expression similar to Eq.(17) with replacing $a-a^+\to a+a^+$ which do not contribute to the gap.

The spin-wave interaction considered in the $1/S$ approximation leads to positive contribution to $\Delta^2$ which is given by 
	\begin{equation}
\Delta^2_{I}=\frac{(A k^2)^2}{4SN}\sum\frac{D_\m q}{D_0},
\end{equation}
and $D_\m q$ is a form-factor of the Dzyaloshinskii interaction \cite{M}, \cite{ER}. The helical structure can be stable if
\begin{eqnarray}
	\Delta^2=\Delta^2_I+\Delta^2_{ME}>0
\end{eqnarray}
and if $\Delta_I=0$ it can survive at $\m k\parallel <100>$ only where $\Delta_{ME}=0$. Meanwhile it is well known that in $Mn Si$ and $Fe Ge$ at low $T$ the helix axis $\m k\parallel <111>$ and $\Delta_I>|\Delta_{ME}|$.  

\section {Experimental consequences  and discussion}
For discussion  experimental consequences of the magneto-elastic interaction we have to know the Young modules $E$ and anisotropic part of the magneto elastic interaction $S^2B_A$. For $Mn Si$ according to \cite{ST}, \cite{GR} the bulk modulus $K=1.37\times 10^{6}bar$ and neglecting the Poisson coefficient $\sigma$ we obtain $E=3K=4,11\times 10^{6}bar$ and $E v_0=240 e V$ ($v_0=95\times 10^{-24}cm^3$).

 Unfortunately the value of $B_A S^2$ is unknown. As we will see below it may be determined by $x$-ray and neutron scattering. Isotropic part of the ME interaction was studied by indirect method in \cite{MA}. Its contribution
   to the lattice constant $\Delta a/a\simeq -1.1\times 10^{-4}$ at $T=0K$  was determined and  the sum of the isotropic part of the ME and elastic energies can be represented as 
\begin{eqnarray}
	B_{is}S^2(\Delta a/a)+K v_0(3\Delta a/a)^2/2.
\end{eqnarray}
This expression is minimal at
 $g_{is}=S^2B_{is}/E v_0=-3.3\times 10^{-4}$ and from Eqs.(15) and (21) we obtain
 \begin{eqnarray}
	E_{ME}&=&-6.5\mu e V\{57m T\}(g_A/g_{is})^2(G_1-G_2/2),\\
	\Delta^2_{ME}&=&-(17\mu e V\{0.15T\})^2(g_A/g_{is})^2(G_1+G_2),
\end{eqnarray}
where we used $S=1.6$, $A k^2=H_C=0.6T$ and $H_C$ is a critical field for transition  to ferromagnetic state \cite{M}.

We compare now  above results with existing experimental data \cite{G3}.  As was shown in \cite{M} the ground state energy of the helical structure in magnetic field is given by
\begin{eqnarray}
	E_G=E_A+E_{ME}-\frac{S H^2_\parallel}{2H_C}-\frac{S H_\perp^2\Delta^2}{4H_C(\Delta^2-H^2_\perp/2)},
\end{eqnarray}
where $H_{\parallel(\perp)}$ is a field component along (perpendicular) to the helix wave-vector $\m k$, $E_A=(S^2F_0 k^2-3S^4K)L/4$,~$F_0$ and $K$ are constants of the anisotropic exchange and cubic anisotropy respectively \cite{B,M}. The cubic invariant $L=(\m g\cdot \hat c)$ is considered in Appendix A.

 Evolution of the helical structure  in magnetic field was studied by small angle polarized neutron scattering in $Mn Si$ near $T_C$ \cite{G2} at low $T$ \cite{G3} and in compound $Fe Co Si$ \cite{G4, G5}. Two new characteristic fields were determined. In zero field the sample is in multidomain state with $\m k$ along all $<111>$ directions. Then  for the field along one of $<111>$ axes at $H_{C1}$  the single domain state appears. With further field increasing the Bragg intensity demonstrated a cusp at $H_{in}$. In \cite{G3} it was interpreted as instability of the $\m k $ direction connected with the second term in Eq.(27). Indeed if $H$ is slightly below $\Delta\sqrt 2$ this term predominates and the vector $\m k$ has to rotate perpendicular to the field but blocked by the anisotropy.  Just below $T_C$ where the anisotropy is weak this rotation was observed in \cite{G2}. In $Mn Si$ we have $H_{C1}=80m T$, $H_{in}160m T$ and $\Delta\simeq 110mT=13\mu e V$ \cite{G3}. 
 
 It is obvious that the single domain state can be realized  if $S H^2_{CI}/2H_C\simeq 9m T$ is of order of $E_A+E_{ME}$. For $<111>$ directions we have $E_{ME}\simeq 25(g_A/g_{is})^2m T$ [See Eq.(A5)]. So this condition is fulfilled roughly. It is impossible to do more detailed analysis as we do not know $E_A$ and $g_A$.
 
    The invariant $L$ has two extrema $L=2/3$ and $L=0$ at $\m k$ along $<111>$ and $<100>$ directions respectively and  a saddle points at $<110>$ directions. Hence if one neglects the ME interaction the configuration with $\m k\parallel <110>$ is forbidden \cite{B}. The same holds for maxima of the critical fluctuations above $T_C\simeq 29K$ \cite{G1}. Meanwhile in $Mn Si$ at high pressure above the quantum critical point $p_c\simeq 14.6 kbar$ maxima of the neutron scattering at $<110>$ directions were observed \cite{P3}. In Appendix B we show that the ME interaction can not resolve this problem i.e. that $<111>$ and $<100>$  remain only possible $\m k$ directions in zero magnetic field.

Let us estimate now  two contribution to the spin-wave gap given by Eqs.(22) and (26). 
  We do not know real form of the  ratio $r=D_\m q/D_0$ in Eq.(22). For $r=1,\:A k^2=0.6T$ and $S=1.6$ we get maximal value $\Delta^2_{I\,max}=(0.24T)^2$.
Minimal value of $\Delta^2_I$ may be estimated assuming that in Eq.(22) $q_{max}=0.024nm$  is a border of the Stoner continuum \cite{ISH}. In this case we have $\Delta^2_{I\, min}=(0.035T)^2$.
In Eq.(26) for $<111>$ direction expression in the bracket is equal to $4/9$ [see Eq.(A5)] and we get $\Delta_{ME}\simeq 0.067(g_A/g_{is})^2T$ . So we see that at ambient pressure  both contributions are comparable with the observed $\Delta\simeq 0.11T$.
Hence we can do plausible assumption that the quantum phase transition observed at $14.6kbar$ \cite{P1, K, P2,P3} is a result of vanishing $\Delta^2$. At higher pressure  the helical structure becomes unstable. 

 For more precise estimations experimental measurement of magneto-elastic anisotropy $S^2(B_1-2B_2)$ would be important. We demonstrate now that it can be  directly extracted from intensities of satellite peaks near nuclear Bragg reflections.  Indeed using Eq.(16) we obtain \cite{NOTE}
\begin{eqnarray}
	\delta I_\pm(\m K)=(2g_A/k)^2| \m K_\pm\cdot[\m g-(\m g\cdot\hat c)\hat c/2]|^2 | F(\m{K_\pm})|^2,
\end{eqnarray}
where $\m{K_\pm=K}\pm 2\m k$, $F(\m Q)$ is the nuclear structure factor,    $\m K$ is a reciprocal lattice point and $g_A=S^2(B_1-2B_2)/E v_0$. Relative satellite intensities are given by
\begin{eqnarray}
	\delta I_\pm(\m K)/I(\m K)\simeq (2g_A/k)^2|\m K\cdot[\m g-(\m g\cdot\hat c)\hat c/2]|^2.
\end{eqnarray}

In  zero magnetic field vectors $\m k$ are along all $<111>$ directions. If  $\m{K\parallel k}$ we have $\delta I_\pm=0$. If however $\m k\parallel (1,1,1)$ but $\m K=(2\pi n/a)(1,1,-1)$ we obtain 
\begin{eqnarray}
	\frac{\delta I_\pm}{I}=\left(\frac{4\pi n g_A}{9\sqrt 3 k a}\right)^2= 2.2\times 10^{-5}\left(\frac{n g_A}{g_{is}}\right)^2
	\end{eqnarray}
	 where we have used $k a=0.17$ \cite{M}. For $\m K\parallel(1,\pm1,0)$ this expression has to be multiplied on $1/4$.

Similar results can be obtained for other $\m K$ directions with one exclusion. We are interested by crystals with $P2_13\:(B20)$ symmetry where $<n,0,0>$ Bragg reflections are forbidden if $n$ is odd and observation of very weak super-lattice reflections would be more easier \cite{CHER}.   Hence this case has to be considered separately. We restrict ourselves by the case $\m k\parallel (1,1,1)$ only. First of all for even $n$ we have  Eq.(30) with replacement $2.2\to 2.2/4\simeq 0.55$. For the odd $n$ the Bragg intensities of the satellites are given by 
\begin{eqnarray}
	I_\pm(\m K)=0.55\times 10^{-5}\left(\frac{n g_A}{g_{is}}\right)^2|F_{Mn}(\m Q)+F_{Si}(\m Q)|^2,
\end{eqnarray}
where form-factors $F_j(\m Q)$ are given in Appendix C. 
They are not zero due $2\m k$ modulation.     Unfortunately  Eq.(31) has an additional small factor  $(2k a)^2<<1$ in comparison with the even-$n$ case. Observation of these odd reflections provides a possibility to determine the lattice chirality [See Eq.(C6)] and its connections with the spin  chirality studied by polarized neutrons \cite{I,G3,G2,G4,G5}. It has to be noted that the lattice chirality in some cases was determined  by anomalous $x$-ray scattering  \cite{I} and electron diffraction \cite{TAN}. 

 There are eight  domains  in virgin sample corresponding $\m k$ along $<111>$   directions. In magnetic field $H_{C1}\simeq 0.08T<<H_C$ the single domain state is realized with $\m k$ along the field and satellite intensity increases. However further increasing of the field suppresses the helical structure and it disappears at $H_C\simeq 0.6T$ \cite{G3}.  In intermediate region at $H_{C1}<H<H_C$  the lattice modulation  with the wave-vector $\m k$ has to appears also \cite{PW}.  Along with discussed  $2\m k$ lattice modulation at low field $H<\Delta\sqrt 2$ the second order helix harmonic appears \cite{M}. I was observed in \cite{G3,LEB}.   

\section{Conclusions}
 We considered the magneto-elastic interaction in cubic helimagnets with $B20$ structure and demonstrated that it deformed the lattice and gave  a negative contribution to the square of the spin-wave gap $\Delta^2$. Hence the helical structure is stabilized due to    positive contribution to $\Delta^2$ which stems from the magnon-magnon interaction \cite{M}. It was suggested that the quantum phase transition observed at pressure in $Mn Si$ and $Fe Ge$ is a result o competition between these two parts of the gap and takes place when $\Delta^2=0$. This suggestion is supported by rough estimations at ambient pressure of both contributions to $\Delta^2$for $Mn Si$ which have the same order and close to experimentally observed gap.
 It was discussed also how to measure directly anisotropic part of the ME interaction responsible for considered phenomena using $x$-ray and neutron scattering.

 \section*{ACKNOWLEDGMENT} 
 The author thanks S.V.Grigoriev and A.V.Syromyatnikov for interesting discussions.
 He is very grateful to S.M.Stishov and D.Yu.Chernyshov also for very important information.
 This work is supported in part by the RFBR (projects No 05-02-19889, 06-02-16702 and 07-02-01318) and the Russian State Programs "Quantum Macrophysics", "Strongly correlated electrons in Semiconductors, Metals, Superconductors and magnetic Materials" and  "Neutron research of solids".

 \appendix
 
 \section{}
 
 In this Appendix we calculate deformation of the lattice by the ME interactions,  cubic invariants $G_{1,2}$ in Eqs.(15), (21), (25-26) and analyse their properties. 
 
 We begin with the classical energy (14).  It is minimal if
\begin{eqnarray}
	w_p+(\m w\cdot\hat c)\hat c_p/(1-2\sigma)=g^*_p,
\end{eqnarray}
where $p=x,y,z$ and $g_p=A^2_p\hat c_p$  and we have
\begin{eqnarray}\nonumber
	w_p&=&g^*_p-(\hat c\cdot \m g^*)\hat c_p/[2(1-\sigma)],\\
	E_{CL}&=&-(2NS^4B^2/Q)[(\m g\cdot\m g^*)-(\m g\cdot\hat c)(\hat c\cdot\m g^*)/(1-\sigma)]
	\end{eqnarray}
	 where $16(\m g\cdot\m g^*)=G_1$ and  $16(\m g\cdot \hat c)(\hat c\cdot\m g^*)=G_2$.

 In cubic $x y z$ frame we can write
\begin{equation}
	\hat a=(\cos\vartheta\cos\varphi,cos\vartheta\sin\varphi,-\sin\vartheta);\:\hat b=(\sin\varphi,-\cos\varphi,0);\:\hat c=(\sin\vartheta\cos\varphi,\sin\vartheta\sin\varphi,\cos\vartheta),
\end{equation}
and for three principal $\m k$-directions $<111>,\:<110>,\mbox{and}\:<100>$ we have: $\{\hat a=(1/\sqrt 6,1/\sqrt 6,-\sqrt{2/3});\:\hat b=(1,-1,0)/\sqrt 2;\:\hat c=(1,1,1)/\sqrt 3\},\:\{\hat a=(0,0,-1);\:\hat b=(1,-1,0)/\sqrt 2;\:\hat c=(1,1,0)/\sqrt 2\},\:\{\hat a=(0,0,-1);\:\hat b=(0,-1,0);\:\hat c(1,0,0)\}$ respectively.
 
In this representation for $G$-functions we obtain
 \begin{equation}
	\begin{aligned}
G_1&=sin^2\vartheta[(\cos^2\vartheta\cos^2\varphi+\sin^2\varphi)^2\cos^2\varphi+(\cos^2\vartheta\sin^2\varphi+\cos^2\varphi)^2\sin^2\varphi+\sin^2\vartheta\cos^2\vartheta]\\&
G_2=\sin^4\vartheta\{[\cos^2\vartheta(1+\sin^4\varphi+\cos^4\varphi)+2\sin^2\varphi\cos^2\varphi]^2+4(\sin^2\varphi-\cos^2\varphi)^2\cos^2\vartheta\sin^2\varphi\cos^2\varphi\}
	\end{aligned}
\end{equation}
From these equations follows that functions $G_1-G_2$ and $G_2$ have extrema at $<111>,\:<110>$ and $<100>$ near which we have
\begin{equation}
	\begin{aligned}
	G_1-G_2&=4/9-20\delta\vartheta^2/9-40\delta\varphi^2/27;\:G_2=8\delta\vartheta^2/9+16\delta\varphi^2/27,\:<111>;\\&
	G_1-G_2=13\delta\vartheta^2/4+\delta\varphi^2;\:G_2=1/4-2(\delta\vartheta^2+\delta\varphi^2),\:<110>;\\&
G_1-G_2=\delta\vartheta^2+\delta\varphi^2;\:G_2=4(\delta\vartheta^4+\delta\varphi^4+\delta\vartheta^2\delta\varphi^2),\:<100>,
	\end{aligned}
\end{equation}
where $\delta\vartheta$ and $\delta\varphi$ are distances from corresponding extremal points. Hence in considered directions both functions $G_1-G_2$ and $G_2$ have extrema and one can show that they have not other extrema.

Contribution of the anisotropic exchange and cubic anisotropy  to the classical energy is proportional to cubic invariant $L$ given by \cite{M}
\begin{eqnarray}
	L=4\sum|A_p|^2\hat c^2_p=  \sin^2\vartheta[(\cos^2\vartheta\cos^2\varphi+\sin^2\varphi)\cos^2\varphi+(\cos^2\vartheta\sin^2\varphi+\cos^2\varphi)\sin^2\varphi+\cos^2\vartheta].
\end{eqnarray}

As above for three principal directions we have
\begin{eqnarray}
	L=2/3-4\delta\vartheta^2/3-8\delta\varphi^2/9,\:<111>;\:1/2+\delta\vartheta^2-2\delta\varphi^2,\:<110>;\:2(\delta\vartheta^2+\delta\varphi^2),\:<100>,
\end{eqnarray}
 and $L$ has a saddle point at $<110>$. 
\section{}
We demonstrate now that in presence  of the ME contribution to the ground state energy the $<111>$ and $<100>$ remain only possible stable directions for the vector $\m k$. 

From Eq.(27) at $\m H=0$ follows
  \begin{equation}
	E_{G}=\Phi L-\Psi(G_1-G_2+G_2/2)=\Psi f(y),   
\end{equation}
 where $\Psi>0$ and $y=\Phi/\Psi$.

We have  to study behavior of $f(y)$ for three principal directions. For $\m k\parallel<111>$ we obtain
\begin{eqnarray}
f(y)=(2/3)(y-2/3)+(4/3)(-y+4/3)(\delta\vartheta^2+2\delta\varphi^2).
\end{eqnarray}
and $E_{G}$ is stable if $\Phi<
 4\Psi/3$. In the $<110>$ case we have
\begin{eqnarray}
f(y)=(1/2)(y-1/4)+(y-9/4)\delta\vartheta^2-2y\delta\varphi^2,
\end{eqnarray}
In this configuration there is a saddle point  as coefficients at deviations $\delta\vartheta^2$ and $\delta\varphi^2$ can not be positive simultaneously.   Finally if $\m k\parallel<100>$ we have
\begin{eqnarray}
	f(y)=2(y-1/2)(\delta\vartheta^2+\delta\varphi^2).
\end{eqnarray}
This configuration is stable if $\Phi>\Psi/2$. However, comparing Eqs.(B2) and (B4) we see that configuration $<111>$ in the region $\Psi/2<\Phi<\Psi/3$   has lower energy and   $<100>$ configuration is metastable. Hence we see that the magneto-elastic energy can not be responsible for the stability of $<110>$ configuration.
 
 \section{}
 
 There are two different ions in compounds with $P2_13$ symmetry ($Mn$ and $Si$ \cite{B,I}; $Fe$ and $Ge$ \cite{LEB} etc.) labeled below as 1 and 2 respectively. Each of them occupy in cubic  unit cell four positions: $\rho_1=(x,x,x),\:\rho_2=(1/2+x,1/2-x,1-x),\:\rho_3=(1-x,1/2+x,1/2-x)\:\mbox{and}\;\rho_4=(1/2-x,1-x,1/2+x)$ (Right-handed structure) or $\rho_1=(x,x,x)\:\rho_2=
(1/2-x,1/2+x,1-x),\:\rho_3=(1/2+x,1-x,1/2-x),\:\mbox{and}\:\rho_4=(1-x,1/2-x,1/2+x)$ (Left-handed structure) \cite{I}.  For $Mn Si$ we have $x_1=0.138$ ($Mn$) and $x_2=0.846$ ($Si$). It is interesting to note that these numbers very close to the ion positions in "ideal" B20 structure with $x_1=1/4\tau=0.1545$ and $x_2=1-x_1=0.8455$ where $\tau=(1+\sqrt 5)/2$ \cite{VOC}.
 
 We consider below the odd-$n$ case only. As we have two different ions in the unit cell the total structure factor is a sum $	F(\m Q)=F_1(\m Q)+F_2(\m Q)$ where
\begin{eqnarray}
	F_j(\m Q)=f_j(\m Q)\sum e^{i(\m Q\cdot\rho_\lambda(x_j))}
\end{eqnarray}
and $f_j(\m Q)$ is a scattering amplitude for the $j$ ion.

 There are eight super-lattice reflections corresponding to $\m Q_1=(2\pi n+\kappa,\kappa,\kappa),\:\m Q_2=(2\pi n-\kappa,\kappa,\kappa),\:\m Q_3=(2\pi n+\kappa,-\kappa,\kappa),\:\mbox{and}\:\m Q_4=(2\pi+\kappa,\kappa,-\kappa)$ where $\kappa=\pm 2k a/\sqrt 3$. Corresponding partial form-factors are given by
\begin{eqnarray}
	F_R(\m Q_1)&=&F_L(\m Q_1)=f e^{2\pi i  n x}[e^{3i\kappa x}-e^{i\kappa(2-x)}]\simeq  i f\kappa (4x-2)e^{2\pi i n x},\\
	F_R(\m Q_2)&=&F_L(\m Q_2)=f\{e^{2\pi i n x}[e^{i\kappa x}-e^{i\kappa(1-3x)}]+e^{-2i\pi n x}[e^{i\kappa x}-e^{i\kappa(1+x)}]\}\simeq i f\kappa [(4 x-1) e^{2\pi i n x}-e^{-2\pi n x}],\\
	F_R(\m Q_3)&=&F_L(\m Q_4)=f\{e^{2\pi i n x}[e^{i\kappa x}-e^{i\kappa(1+x)}]+e^{-2\pi i n x}[e^{i\kappa(1-3x)}-e^{i\kappa x}]\}\simeq i f\kappa [-e^{2\pi i n x}+(1-4x)e^{-2\pi i n x}],\\
	F_R(\m Q_4)&=&F_L(\m Q_3)=f e^{-2\pi i n x}[e^{i\kappa(1+x)}-e^{i\kappa(1-3x)}]\simeq 4f i \kappa x e^{-2\pi i n x}.
\end{eqnarray}

From Eqs.(C4,5) follow the intensity ratio
\begin{eqnarray}
	\frac{I_R(\m Q_3)}{I_R(\m Q_4)}=\frac{I_L(\m Q_4)}{I_L(\m Q_3)},
\end{eqnarray}
and from measurement of the $2\m k$ lattice deformation one can determines the lattice chirality.


\begin{thebibliography}{99}

\bibitem{HK} H.Katsura, A.V.Balatsky, N.Nagaosa,
	Phys.Rev.Lett. \textbf{98}, 027203 (2007).
\bibitem{D} I.E.Dzyaloshinskii, Zh.Eksp. Teor.Fiz.\textbf{46}, 1420 (1964)[Sov.Phys.JETP\textbf{19}, 960 (1964)]; \textbf{47},336 (1964)[\textbf{20}, 223 (1965)]; \textbf{47}, 992 (1964)[\textbf{20}, 665 (1965)].
\bibitem{N} O.Nakanishi, A.Yanase, A.Hasegava, M.Kataoka, Solid State Commun.\textbf{35}, 995 (1980).
\bibitem{B} P.Bak, M.Jensen, J.Phys. \textbf{C 13}, L881 (1980). 
\bibitem{I} M.Ishida, Y.Endoh, S.Mitsuda, Y.Ishikawa, M Tanaka,J.Phys.Soc.Jpn. \textbf{54} 2975 (1975).
\bibitem{P1} C.Pfleiderer, G.J.MacMillan, S.R.Julian, G.G.Lonzarich, Phys.Rev. \textbf{B 55}, 8330 (1997).
\bibitem{K} K.Koyama, T.Goto, T.Kanomata, R.Note, Phys.Rev. \textbf{B 62},986 (2000).
\bibitem{P2} C.Pfleiderer, S.R.Julian, G.G.Lonzarich, Nature(London) \textbf{414}, 427 (2001).
\bibitem{P4} C.Pfleiderer, D.Reznik, L.Pintschovius, and J.Haug, Phys.Rev.Lett. \textbf{99}, 156406 (2007).
\bibitem{PE} P.Pedrazzini, H.Wilhelm, D.Jaccard,T.Jalborg,M.Schmidt, M. Hanfland, l. Akselrud, H.Q.Yuan, U.Schwarz, Yu.Grin, and F.Steglich, 	Phys.Rev.Lett., \textbf{98}, 047204 (2007) 
\bibitem{P3} C.Pfleiderer, D.Reznik, L.Pintschovius, H.v.L\"ohneysen, M.Garst, A.Rosh, Nature (London)\textbf{427}, 227 (2004). 
\bibitem{T} S.Tewari, D.Belitz, T.R.Kirpatrick, Phys.Rev.Lett.\textbf{96}, 047207 (2006).
\bibitem{R} U.K.R\"ossler, A.N.Bogdanov, C.Pfleiderer, Nature(London) \textbf{442}, 797 (2006).
\bibitem{BK} D.Belitz, T.R.Kirpatrick, A.Roch, Phys.Rev. \textbf{B 73}, 054431 (2006).
\bibitem{Bn} B.Binz, A.Vishwanath, Phys.Rev. \textbf{B 74}, 214408 (2006).
\bibitem{G1} S.V.Grigoriev, S.V.Maleyev, A.I.Okorokov, Yu.O.Chetverikov, R.Georgii, P.B\"oni, D.Lamago, H.Eckerlebe, K.Pranzas, Phys.Rev. \textbf{B 72},134420
(2005).
\bibitem{ST}S.M.Stishov, A.B.Petrova, S.Khasanov, G.Kh.Panova, A.A.Shikov, J.C.Lashley, D.Wu, and T.A.Logasso, Phys.Rev. \textbf{B 76}, 052405 (2007).
\bibitem{U} Y.J.Uemura. T.Goko, I.M.Gat.Malureanu et al., Nature Physics \textbf{3}, 29 (2007).

  
 \bibitem{M} S.V.Maleyev, Phys.Rev. \textbf{B 73}, 174403 (2006).
  
 \bibitem{M1} S.V.Maleyev, arXiv: 0711.3547.
 \bibitem{com} Rotation two different spins on infinitesimal angle $\vec \varphi\parallel \m k$ does not change  the exchange energy  but add $2D\sum_{\m{1,2}}(\nabla_1-\nabla_2)[\m S_1(\vec \varphi\cdot\m S_2)-\m S_2(\vec\varphi\cdot\m S_1)]=2D i\sum[\m{(q\cdot S_q)}(\vec\varphi\cdot \m{S_{-q}})-(\vec\varphi\cdot \m{S_q)(q\cdot S_{-q}})]\neq 0$ to the Dzyaloshinskii interaction.
 
 \bibitem{PW1} M.L.Plumer and M.B.Walker, J.Phys.C: Solid State Phys. \textbf{14}, 4689 (1981).
 \bibitem{G3}S.V.Grigoriev, S.V.Maleyev, A.I.Okorokov,  Yu.O.Chetverikov, P.B\"oni, R.Georgii, D.Lamago, H.Eckerslebe and K.Pranzas, Phys.Rev. \textbf{B 74},214414 (2006).
 \bibitem{G2} S.V.Grigoriev, S.V.Maleyev, A.I.Okorokov, Yu.O.Chetverikov, and H.Eckerlebe, Phys.Rev. \textbf{B 73},224440 (2006).
 
\bibitem{G4} S.V.Grigoriev, S.V.Maleyev, V.A.Dyadkin, D.Menzel, J.Shoenes, and H.Eckerlebe, Phys.Pev. \textbf{B 76},092407 (2007).
\bibitem{G5} S.V.Grigoriev, V.A.Dyadkin, J.Schoenes, Yu.O.Chetverikov, A.I.Okorokov, H.Eckerlebe, and S.V.Maleyev, Phys.Rev. \textbf{B 76}, 224424 (2007).
 
 
 
 
\bibitem{GE} G.A. Gehring, arXiv:0711.2586.
 
\bibitem{PW} M.L.Plumer and M.B.Walker, J.Phys.C: Solid. State Phys.\textbf{15},7181 (1982). 
 \bibitem{P}M.L.Plumer, J.Phys.C: Solid.State.Phys., \textbf{17}, 4663 (1984).
 
 
 \bibitem{L} L.D.Landau and E.M.Lifshitz, Electrodynamics of Continuous Media (Pergamon Press, Oxford, 1984).
     \bibitem{LL} L.D.Landau and E.M.Lifshitz, Statistical Physics, Pt 1 (Pergamon Press, Oxford,1969).

 \bibitem{Ref} As in \cite{M} we use term umklapp for processes mixing excitations with momenta $\m k$ and $\pm\m k,\:\pm 2\m k$ etc..
 \bibitem{NB} They are allowed in magnetized state only \cite{PW}.  
 \bibitem{LL1}L.D.Landau and E.M.Lifshitz, Theory of Elasticity (Pergamon Press, Oxford, 1986). 
\bibitem{ER}At $H=0$ cubic anisotropy does not contribute to $\Delta^2$ and  Eqs.(52-53) in \cite{M} are erroneous.



  


\bibitem{GR} E.L.Gromnitskaya,  Ref.[23] in\cite{ST}. 
\bibitem{MA} M.Matsunaga, Y.Ishikawa, and T.Nakajima, J.Phys.Soc.Jpn. \textbf{51}, 1153 (1982).
\bibitem{ISH}Y.Ishikawa, G.Shirane, J.A.Tarvin, and M.Kohgi, Phys.Rev. \textbf{B 16}, 4956 (1977).
\bibitem{NOTE} Equations presented below are correct if   the   lattice and the helical structure have the same  mosaic.  For $Mn Si$ the magnetic mosaic is greater than the lattice one \cite{G3} and an additional small factor has to be introduced in expressions for the relative intensities.  
\bibitem{CHER} I am thankful D.Yu.Cheryshov for  corresponding explanation.
\bibitem{TAN} M.Tanaka, H.Takajoshi, M.Ishida, and Y.Endoh, J.phys. Soc.Jpn., \textbf{54}, 2970 (1985). 

 
\bibitem{LEB} B.Lebech, J.Bernard, and T.Freltoft, J.Phys.: Condens.Matter \textbf{1}, 6105 (1989).
\bibitem{VOC} L.Vo\u{c}adlo, G.D.Price, and I.G.Wood, Acta.Cryst., \textbf{B 55}, 484 (1999).

\end{thebibliography}
\end{document}